\documentclass[aps,prl,twocolumn,showpacs]{revtex4}
\usepackage{amsmath}
\usepackage{amssymb}
\usepackage{graphicx}
\usepackage{color}

\begin{document}

\title{Mechanical Bounds to Transcriptional Noise}
\author{Stuart A. Sevier,$^*$ }
\affiliation{Department of Physics and Astronomy, Center for Theoretical Biological Physics, Rice University, Houston, TX 77005, U.S.A.}

\author{David A. Kessler}
\affiliation{Department of Physics, Bar-Ilan University, Ramat-Gan IL52900, Israel}
\date{\today}

\author{Herbert Levine }
\affiliation{Department of Bioengineering, Center for Theoretical Biological Physics, Rice University, Houston, TX 77005, U.S.A.}
\date{\today}

\begin{abstract}

Over the last several decades it has been increasingly recognized that stochastic processes play a central role in transcription.  Though many stochastic effects have been explained, the source of transcriptional bursting (one of the most well-known sources of stochasticity) has continued to evade understanding. Recent results have pointed to mechanical feedback as the source of transcriptional bursting but a reconciliation of this perspective with preexisting views of transcriptional regulation is lacking. In this letter we present a simple phenomenological model which is able to incorporate the traditional view of gene expression within a framework with mechanical limits to transcription.  Our model explains the emergence of universal properties of gene expression, wherein the lower limit of intrinsic noise necessarily rises with mean expression level.

\end{abstract}

\maketitle

The ability to watch biological phenomena play out at the single molecule level has revealed a rich and nuanced view of the central dogma of biology.  From the single molecule vantage it has become clear that random forces and events play a key role in transcription  \cite{Raj2008}. The identification of transcriptional bursting, in which genes undergo periods of paused activity even in fully induced environments \cite{Golding2005}, has been one of the most notable examples of this new perspective. Bursting has also figured prominently in the discussion concerning universal properties of transcriptional noise \cite{Sanchez2013}. In particular, a number of recent experimental results have found that there is a link between the rate and randomness of mRNA production where highly expressed genes have increased noise associated with production \cite{Sanchez2013}. This result transcends specific organisms or genes, and may be explained if expression inevitably exhibits bursting. Other work, however, has argued that under some conditions there are non-universal gene specific relationships between the rate and randomness of mRNA production \cite{Jones2014}; these results are more consistent with the pure model  of transcription regulated by the binding of specific regulatory proteins to the promoter regions, in which high expression is not necessarily associated with high noise.

 \begin{figure}[t]
\includegraphics[width=0.9\linewidth]{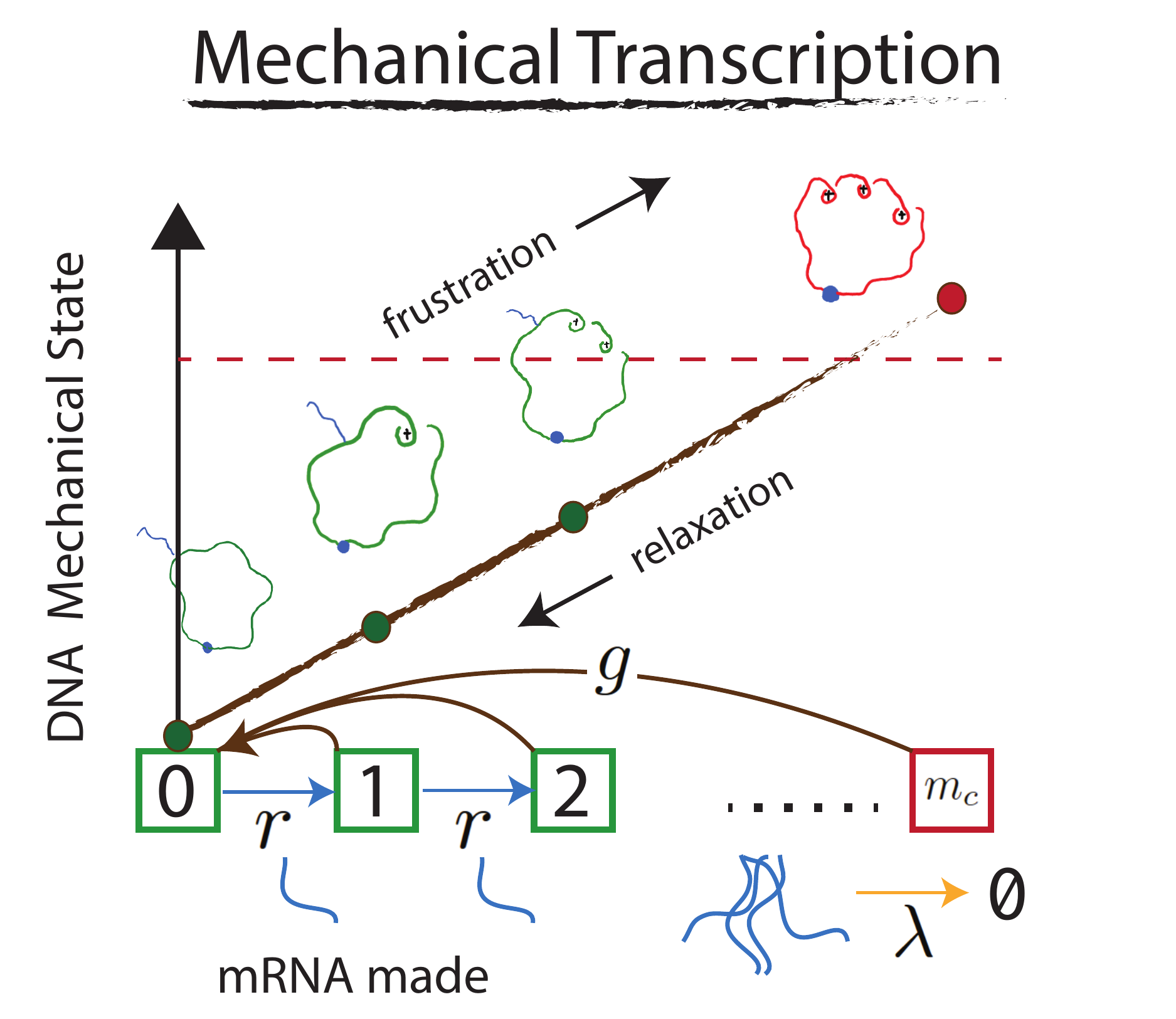}
 \caption{(color online) A cartoon depicting the relationship between transcriptional dynamics and mechanical memory.  As transcripts are made at rate $r$, the mechanical state increases in strain associated with DNA winding, leading to a mechanical limitation to transcription prior to a relaxation step, occurring at rate $g$. }
 \end{figure}

What is needed is a framework which is able to accommodate both the traditional `promoter architecture' view of transcription while at the same time capturing recently observed universal aspects of bursting. To accomplish this, we start from  the ``twin-supercoiling domain" \cite{Liu1987} model of transcription wherein the helical nature of DNA combined with topological obstructions leads to the accumulation of mechanical strain in DNA during transcription.  This strain can result in arrested gene expression.  Specific biological machinery (topoisomerases) must relieve the strain created by transcription through physical, topological manipulation of the DNA in order for gene expression to continue. A recent study has shed further light on these mechanical aspects of transcription covering both the physical range and speed at which RNA polymerase (RNAP) can operate \cite{Ma2013}.  Additionally the self-induced stalling and topoisomerase mediated recovery of RNAP during transcription in bacteria \cite{Chong2014} has been observed in real-time, highlighting the intrinsic role of super-coiling and mechanics in gene expression at the single transcript level. 

  \begin{figure}[t]
 \includegraphics[width=0.9\linewidth,clip=]{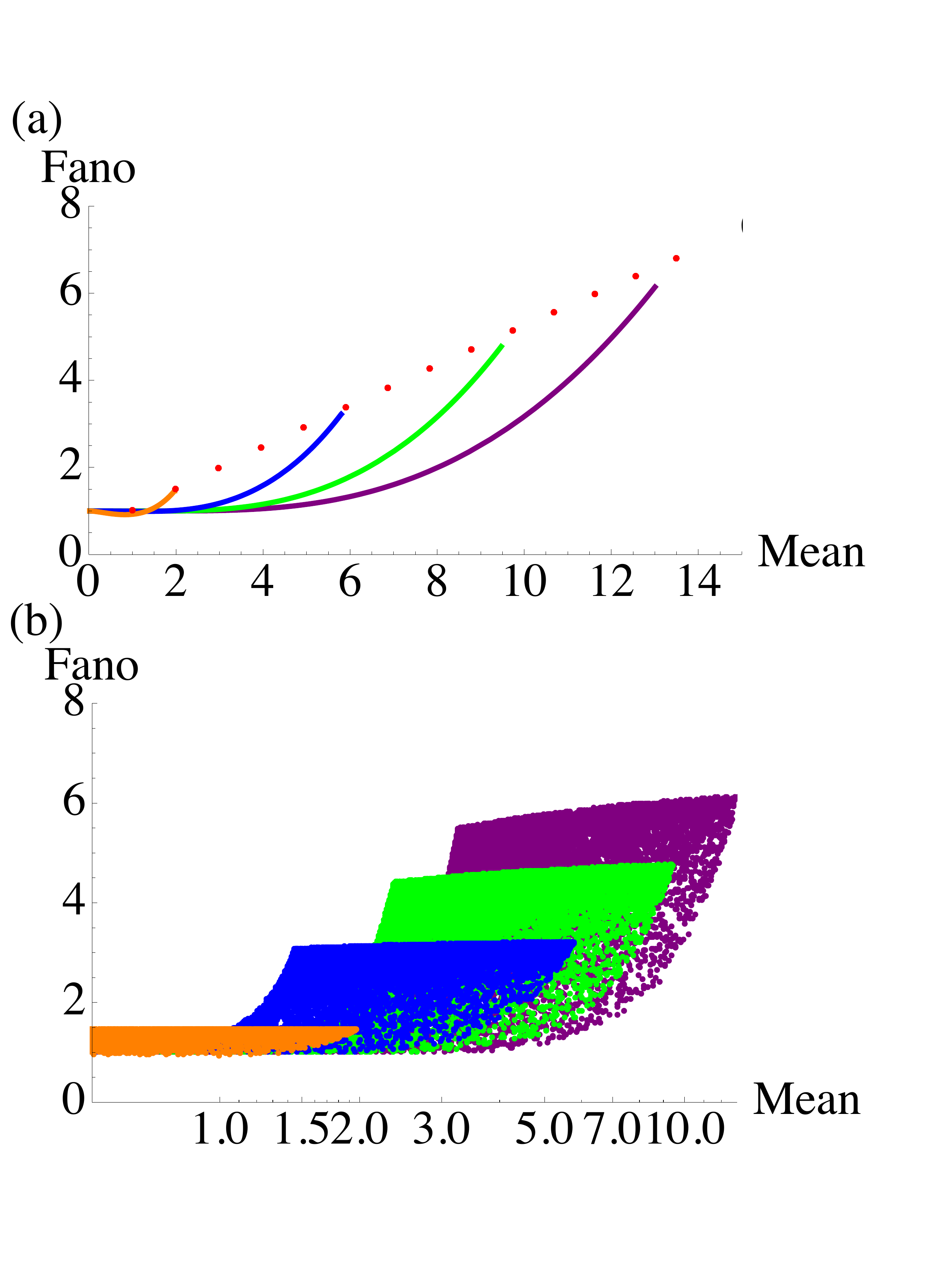}
 \caption{(color online)(a) Curves showing the relationship between mean mRNA and  noise (Fano factor)  for increasing transcriptional rate, for various cutoff values $m_{c}=2,6,10,14$ at fixed $\lambda=\frac{1}{20},\;g=\frac{1}{20}$. The infinitely fast transcriptional limit is shown by red dots. (b) Macroscopic exploration of possible noise (Fano factor) mean relationships for a variety of transcriptional rates $r\in[0,5]$ and decay rates $\lambda \in[ \frac{1}{20}, \frac{1}{5}]$ }
 \end{figure}

Such a mechanically based regulatory scheme acts not independently of, but instead underneath, well-known biological regulatory agents, calling for an extension of the standard view of gene regulation.  We believe that this perspective is crucial since mechanical silencing requires transcription so that in systems with mechanical stalling the rate of mechanical arrest is fundamentally tied to the rate of mRNA production. 

A simple way to incorporate the points outlined above is the introduction of a hard limit to the number of transcripts any particular gene is able to make before expression arrests and must thereafter wait for a randomly occurring relaxation event before it can resume. We therefore propose the following stochastic model. Each mRNA transcript is made with rate $r$ and decays independently with rate $\lambda$ .  The rate of mRNA production $r$ reflects the ability of a particular gene to promote transcription.  As each mRNA is made the mechanical strain inside the DNA is increased until RNAP can no longer operate, which corresponds to a deterministic maximum number of transcripts $m_{c}$ being made (and a corresponding mechanical state of the DNA begin reached)  (depicted in Fig.1).   Relaxation events occur with rate $g$, relieving all mechanical strain in the DNA and erasing the gene's  knowledge of previous transcripts made. A pivotal assumption is that for each transcription event $m\rightarrow m+1$ the internal mechanical strain of the DNA increases by some amount which must be removed by a subsequent relaxation event.  This makes the non-equilibrium cycle of frustration and relaxation a fundamental part of transcription. 
We first examine the simpler case of  pure mechanical regulation before incorporating this mechanism into a standard repressor regulatory scheme.

\begin{figure*}[t]
\includegraphics[width=0.9\linewidth,clip=]{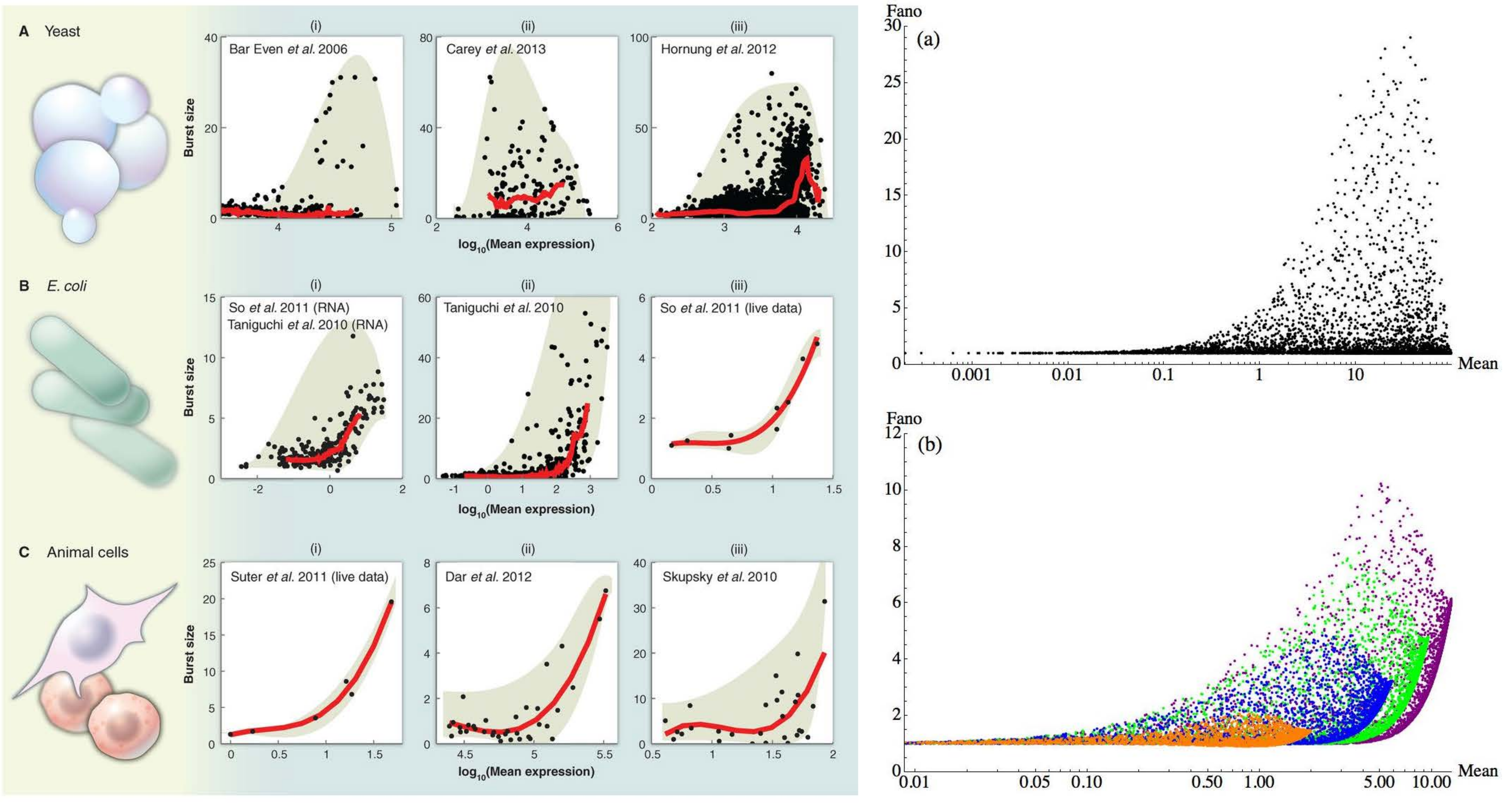}
 \caption{(color online)A comparison of experimental (left) noise/mean relationships (Fig.2 from \cite{Sanchez2013}) to results obtained in this letter (right) for a gene with both mechanical and repressive regulation. Note the characteristic rise in Burst size (Fano factor) with expression level in the experimental data for {\em E. coli} and animal cells (but not for yeast). Theoretical plots are explorations of possible noise/mean relationships for a variety of transcriptional rates $r\in[0,5]$  and repressor rates $k_{\pm}\in[\frac{1}{100},10]$ for fixed $\lambda=\frac{1}{20},\;g=\frac{1}{20}$.  Right (a) Colors correspond to different cutoff values $m_{c}=2,6,10,14$  and the lower bound to noise values and limited production strength closely resembles rows B and C of Fig. 2 from \cite{Sanchez2013}. Right (b) no mechanical limit is present and there is no lower bound to noise values at any mean expression level closely resembling row A of Fig. 2 from \cite{Sanchez2013}.}
\end{figure*}

The most natural mathematical framework in which to incorporate all the elements described above is the master equation through which stationary expressions for the mean and variance of mRNA levels can all be analytically derived (see supplementary material for details).  As a transcript is made both the number of transcripts present, $m$,  as well the internal strain of the system $P_{\alpha}(m)\rightarrow P_{\alpha+1}(m+1)$ increase, where $\alpha$ denotes the number of transcripts made since the last relaxation event. At the cutoff, $\alpha=m_{c}$, transcription stalls until a relaxation event resets the system to $\alpha=0$. The probability of finding $m$ mRNAs is the sum over each internal mechanical state $P(m)=\sum_{\alpha} P_{\alpha}(m)$.  The moments of this probability distribution are most easily obtained by introducing a generating function $G_{\alpha}$ for each $P_{\alpha}$.  By doing this the equation for the  total generating function $G(z)=\sum_{\alpha}G_{\alpha}(z)$ only depends on the arrested state, viz.,
\begin{equation}
   \dot{G}= \lambda(1-z)\partial_{z}G+r(z-1)(G-G_{{m_c}})
\end{equation}
By setting $ \dot{G}=0$ and taking various partial derivatives of $G$ we can calculate the stationary moments of the mRNA levels.  The only challenge exists in calculating $P_{{m_c}}$ and $\overline{m}_{m_c}$ the probability of being in the arrested state and the expected number of mRNA given that the system is in the arrested state, respectively.  Both can be obtained analytically and yield an occupational probability and mean mRNA expression level which are non-linear in the transcriptional and relaxation rates 
\begin{equation}
 \overline{m}=\frac{r}{\lambda}\left(1-P_{{m_c}}\right),\; P_{m_c}=\left(\frac{r}{r+g}\right)^{m_{c}}
\end{equation}
One simple limit is where transcription occurs very quickly compared to the other time-scales.  One can treat this case as a simple  process of making a burst of $m_c$ transcripts at rate $g$, plus mRNA decay at rate $\lambda$. This leads to the mRNA level becoming independent of the transcriptional rate, $\overline{m}\rightarrow m_{c}\frac{g}{\lambda}$. Similarly, the noise in this limit, as captured by the Fano factor $F=\frac{variance(m)}{mean(m)}$, is  dependent only on the cutoff value, $F \rightarrow \frac{1+m_{c}}{2}$.  In general, the expression for the Fano factor is complicated, and is given in the SM.

The relationship between mean and noise levels as a function of promoter strength $r$ for various cutoff values at fixed relaxation rate $g$ is shown by the curves in Figure 2a with the red dots showing the linear dependence in the large $r$ limit. Of course, at small $r$ the system never reaches the cutoff state and the process is purely Poissonian, with $F=1$. The Fano factor begins to rise towards the aforementioned asymptotic value when the time to mechanical frustration becomes comparable to the time of relaxation, so that $\frac{r}{m_{c}}\gtrsim g$.  A useful approximation to the curve can be obtained by going to the limit of large $r$, large $m_c$, keeping  the ratio $x=\frac{r}{m_{c}g}$ fixed.  For the special case of $g=\lambda$ (as in Figure 2a), the curve in this scaling limit takes the form
\begin{equation}
 \overline{m}\approx m_{c} x \left(1-e^{-\frac{1}{x}}\right),\;F\approx m_{c}\frac{xe^{-\frac{1}{x}}(1-e^{-\frac{1}{x}})}{2}
\end{equation}
This result, which incorporates the above large $r$ limit, illustrates how a mechanical limit to transcription provides a natural mechanism for bounding from below the intrinsic noise at a given mean expression level and thus demonstrates how mechanical regulation provides an explanation of this universal property of gene expression noise. 

To see how large-scale transcriptional data might appear when different genes are characterized by a spectrum of rates as might exist in real organisms, we have taken the analytical expressions for the mean mRNA levels and Fano factors and generated theoretical data clouds by varying transcriptional  production and decay rates $r,\lambda$, all for different cutoff levels $m_{c}$, in Figure 2b.    A change in the mechanical barrier through the cutoff value $m_{c}$ postpones the rise in the Fano factor to larger mean expression level, and so decreases the lower bound of the cloud.  At the same time, it increases the maximum Fano factor.   We see that the presented clouds show that the minimum Fano factor increases with increasing mean expression. Thus, the introduction of mechanical limitations to transcription provides an explanation for the previously unexplained high noise - high  mean properties of gene expression \cite{Sanchez2013,So2011} and clearly illustrates how macroscopic bounds of gene expression data might arise.   Increasing the cutoff value $m_{c}$ or increasing the rate of mechanical relaxation decreases the lower bound on transcriptional noise and offers a plausible explanation for recent observations of weak noise for constitutively transcribed genes \cite{Jones2014}. A  fuller understanding of the mechanical properties of any particular gene is needed to make more direct comparisons. 

Most genes are subject to regulation by proteins that couple to DNA and affect RNAP recruitment. For concreteness, we will focus on the case of a repressor. To include well-known repressor effects, we simply posit that the production of mRNA will be paused while the gene is repressed.   We introduce a hypothetical repressor with binding/unbinding rates $k_{\pm}$ corresponding to states in which the previously described mechanical process is active (repressor unbound) or paused (repressor bound).  While active, the gene is governed by the same underlying mechanical process as previously described; and since the promoter-based silencing is independent of the topoisomerase action, we will allow for relaxation events to take place for both active and repressed genes.

Using the same master equation approach as in the unregulated case we can derive analytical results for the stationary mRNA mean level 
\begin{equation}
\overline{m}=\frac{r}{\lambda}P_\textit{on}\left(1-P_{m_c}\right)\end{equation}
\begin{equation}
P_\textit{on}=\frac{k_{-}}{k_{-}+k_{+}},\; P_{m_c}=\left(\frac{r}{r+g+g\frac{k_{+}}{g \ + \ k_{-}}}\right)^{m_{c}}
\end{equation}
In the limit where the mRNA production rate $r$ is the fastest rate present we again find an expression for the stationary mean which is $r$ independent $\overline{m}\rightarrow m_{c}\frac{g}{\lambda}P_{on}(1+\frac{k_{+}}{g \  + \ k_{-}})$ as well as an $r$ independent noise level. For repressor dynamics which is much slower than the relaxation rate, $k_{\pm}\ll g$, we have

\begin{equation}
F\sim\frac{1+m_{c}}{2}+m_{c}\frac{gk_{+}}{(\lambda+k_{+}+k_{-})(k_{+}+k_{-})}
\end{equation}
This formula is very instructive. The Fano factor is composed of two contributions. The first is that due to the mechanical relaxation alone.  The second is exactly the Fano factor for the pure regulatory dynamics, with $(r/\lambda)P_\textit{on}$ replaced by $(g m_c/\lambda)P_\textit{on}$, which is the mean expression level at large $r$ in the presence of mechanical regulation.  In particular, the formula illustrates how the noise can be very large if $k_{+}> k_{-}$ and $g>\lambda$, explaining points in the data which are much greater than the  noise bound set by mechanical regulation. As expected, the repressor diminishes the mean level of expression and also has significant effects on the intrinsic noise. 

 \begin{figure}[t]
\includegraphics[width=0.9\linewidth,clip=]{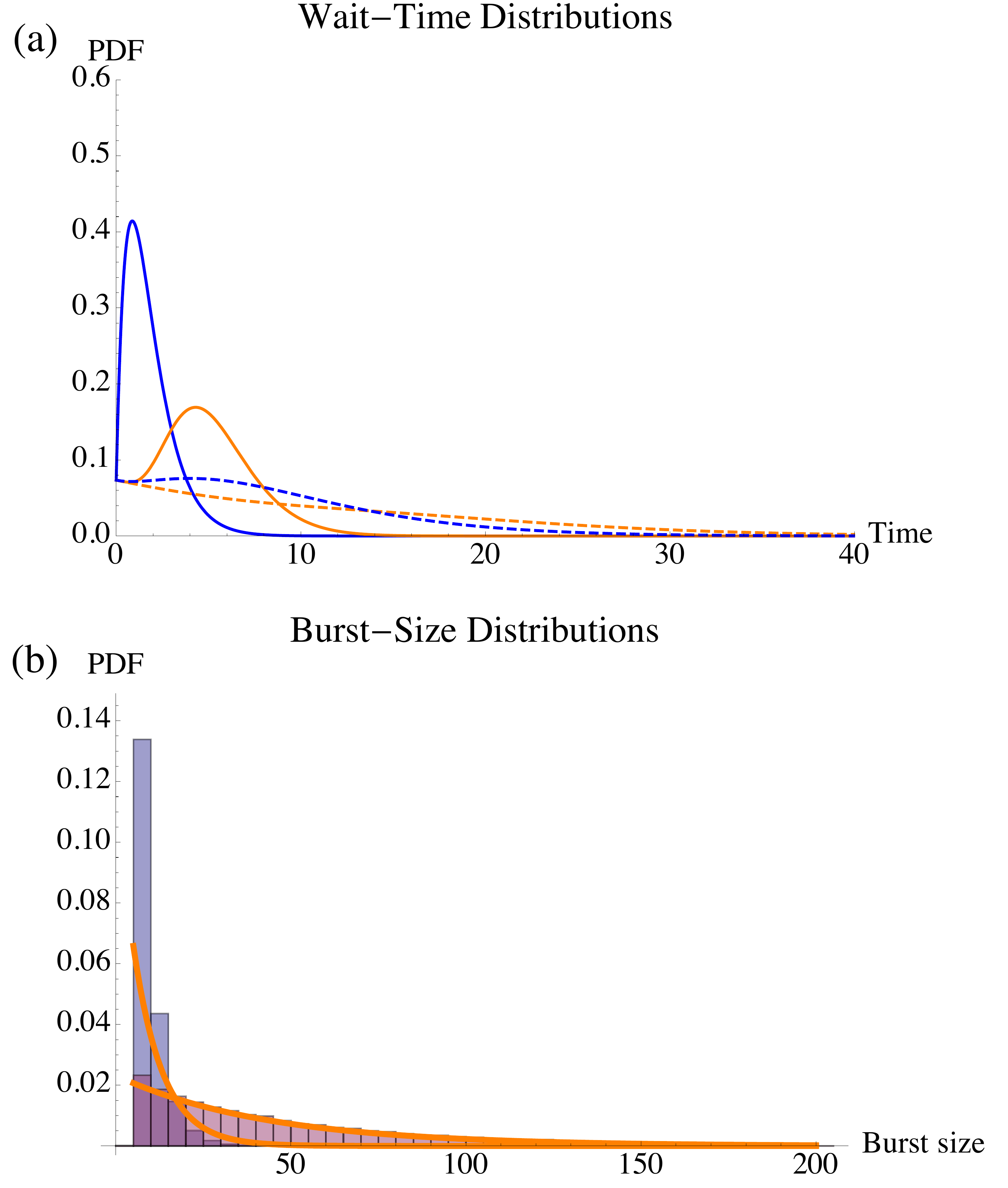}
 \caption{(color online) (a) Wait-Time probability distributions for cutoff values $m_{c}=2,6$ (blue, orange) for $r=1,\frac{1}{2}$(solid, dashed) at fixed $g=\frac{1}{20},\;k_{+}=\frac{1}{50}$ (b) Burst-size distributions showing the histograms of the number of mRNA produced before arresting for two relaxation rates for fixed  $r=1,\;m_{c}=5$.  Overlaid orange curves show best fits to geometric distributions.}
 \end{figure}
 
In addition to changing expression levels by varying $r$, we can vary the  repressor binding/unbinding rate $k_{\pm}$ ratio for the case of a repressed gene.   Interestingly, the noise/mean curves for a repressed gene show decreased noise (at a fixed mean) with mechanical governance as opposed to a non-mechanical repressed gene, as this ratio is varied (see Fig. S5).  The noise/mean relationship for varying repressor kinetics produces curves similar to those seen in recent experiments \cite{Jones2014}.

We can again explore all possible noise/mean relationships for a multitude of hypothetical genes faced with mechanical limits and make a direct comparison to previously published universal gene expression noise/mean data \cite{Sanchez2013}. The theoretical data points in Figure 3 represent mean/noise values for mRNA being produced and decaying for a variety of rates $r,k_{\pm}$ at fixed relaxation and decay $g,\lambda$. The signature rise in the lower bound of the noise with increased expression that we encountered for  non-repressed genes is preserved and it is again shifted by changing the mechanical barrier through the cutoff value $m_{c}$. This is most clearly illustrated by contrasting the colored points of Figure 3, which are subject to mechanical limits, to the black points which are not. The introduction of a mechanical limit to transcription naturally generates  previously unexplained bounds to the  mean and intrinsic noise found in experimental data \cite{So2011, Sanchez2013}.
 
Beyond noise/mean statistical data, wait-time distributions for periods of activity and inactivity, as well as the burst-size distributions for the number of mRNA made in-between arrests, can shed light on the nature of the arrest process. We operationally defined bursts as periods without mRNA production lasting more than a time $\frac{2}{\lambda}$ while wait-times represent the time to reach the arrested state from the completely relaxed state. In the case where $r\gg g$, the gene never relaxes before arresting and the wait-time distributions have short-time peaks (see solid curves in Figure 4a) but exponential long-time behavior as well as a simple burst-size distribution where the maximum number of transcripts $m_{c}$ are typically made between arrests (see purple histogram of Figure 4b).  However, when the transcription rate is more comparable to relaxation, the peaks of the wait-time distributions become broadened and the distributions show strong exponential behavior (see dashed curves in Figure 4a).  Additionally, for slowly arresting genes, a much broader burst-size distribution, where often many more than $m_{c}$ transcripts are made in between arrest events, is observed (see blue histogram of Figure 4b) showing a strong resemblance to a geometric distribution.

It is worth noting that the true nature of the mechanical arrest problem, which takes into account the physical drag and energy associated with the act of transcription as well as the precise action of the topoisomerases, is not addressed in detail within this letter.  However, the simple phenomenological model presented here is a first attempt at putting together the necessary ingredients for understanding this more detailed problem and appears to capture several essential features of transcriptional noise.  As a way of including a more nuanced arrest mechanism we have conducted numerical simulations (shown in Figure 7 as points) for a gene which experiences decreased proclivity for expression as the cutoff $m_{c}$ is approached and found that only slight quantitative changes occurred (see S.M. for details). We expect further theoretical and experimental work  on the role mechanical effects play in gene expression to refine, but not significantly change, the perspective presented here.   

Finally, the theoretical discussion within this letter has not centered on any particular organism.  There is no reason to believe that the framework constructed here is not capable of capturing the same phenomena in many organisms, offering an explanation for the noise/mean relationship observed in both bacteria and higher organisms \cite{Sanchez2013}.  Though the most direct evidence for transcriptional bursting and supercoiling based arrest exists in bacterial systems, the first discovery of topoisomerases and transcription induced supercoiling was done in eukaryotes  \cite{Wang2002}.  Additionally in yeast it has been demonstrated that transcription can occur without topoisomerases \cite{Wang2002}  possibly due to reduced chromosome organization \cite{Duan2010}.  Thus mechanically induced stalling provides a mechanism through which previously unexplained mean and noise bounds in transcriptional data across many organisms \cite{Sanchez2013} can be reconciled with the prevailing 'promoter architecture' view of transcription.   

In conclusion the results of this letter are the first steps in combining the traditional view of transcriptional regulation with recent discoveries concerning the role of stochasticity and mechanics.  The reconciliation of these two perspectives, even at the simple level presented here, is key to resolving outstanding puzzles in gene expression and allowing for a more complete view of transcription to emerge. 

\acknowledgments {We thank Samuel Skinner and Ido Golding for helpful discussions. This work was supported by the National Science Foundation Center for Theoretical Biological Physics (Grant NSF PHY-1308264). HL  was also supported by the CPRIT Scholar program of the State of Texas.  DAK was supported by the Israel Science Foundation (Grant 376/12).}

\bibliographystyle{unsrt}
\bibliography{mechanical4_final}

\pagebreak

\onecolumngrid
\appendix

\section{Supplementary Material for ``Mechanical Bounds to Transcriptional
Noise''}

\begin{figure}[!h]
\begin{center}
\includegraphics[width=0.7\textwidth]{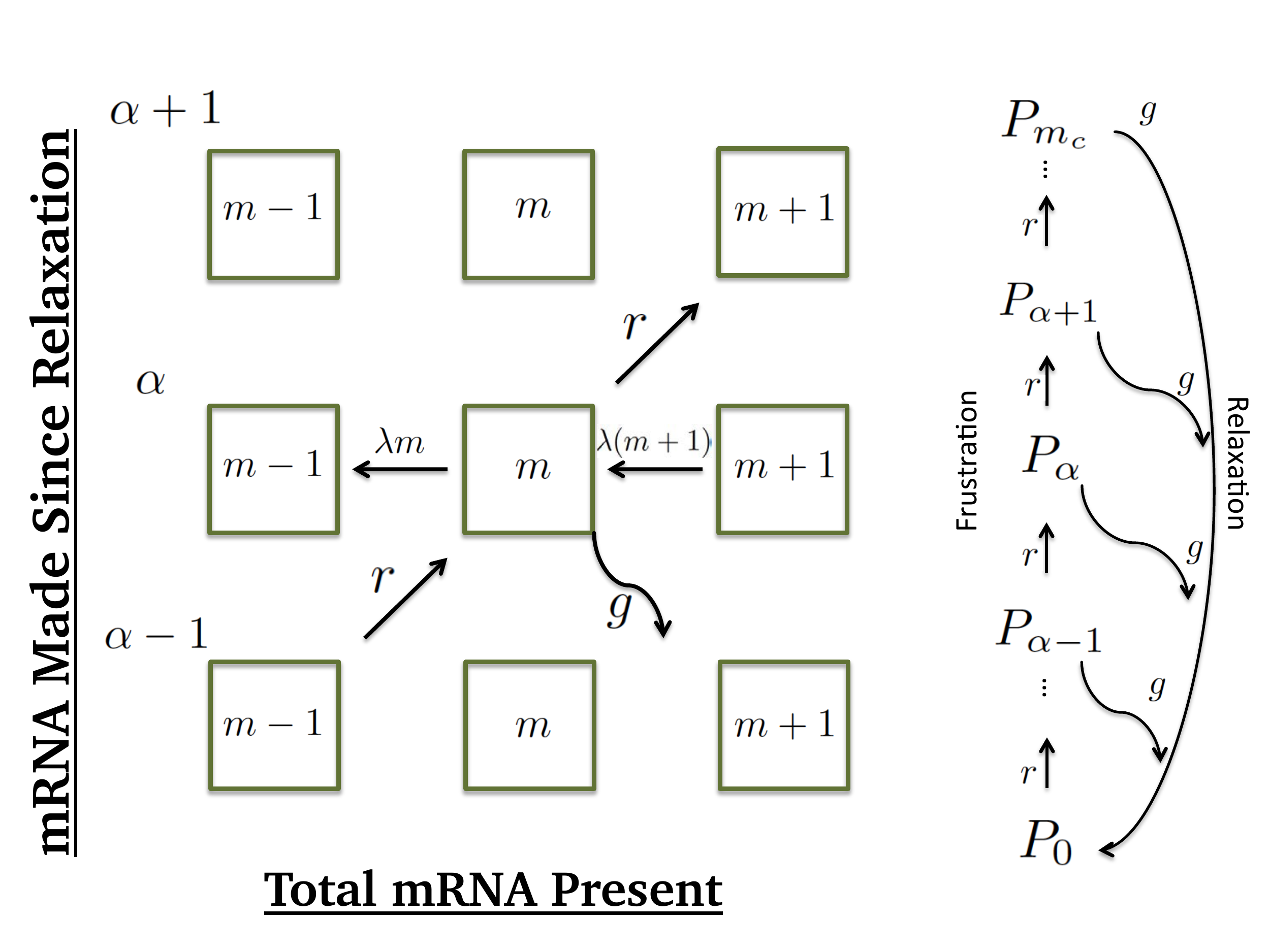}
\end{center}
\caption{
A cartoon depicting the structure of the states within the
system.}
\label{fig2}
\end{figure}

\section{Pure Mechanical Model: Generating Function Approach}

We are going to construct a master equation for the state probabilities $P_\alpha(m)$, $\alpha=0,1,\ldots,m_c$, $m=0,1,\ldots$.  The state space and transitions are illustrated in Fig. \ref{fig2}.  From this, we obtain equations for the generating function for each state $\{P_{\alpha}\}$
as $G_{\alpha}(z,t)=\sum_{m}z^{m}P_{\alpha}(m,t)$.  From the $G_\alpha$ we can construct
the entire generating function as $G(z,t)=\sum_{i}G_{i}=\sum_{\alpha}\sum_{m}z^{m}P_{\alpha}(m,t)=\sum_{m}z^{m}P(m,t)$
where $P(m,t)$ is the probability of $m$ mRNAs, traced over the mechanical states $\alpha$. We will proceed by writing
down three separate master equations which correspond to the three
distinct types of states in the system and for each, transform them
into equations for the respective generating functions from which
we can calculate the zeroth moments (traced over $m$, for a given $\alpha$), namely $P_\alpha$.

\subsection{Equations}

\subsection*{Relaxed ($\bm{\alpha=0}$)}

\begin{equation}
\dot{P}_{0}  =  -(r+g+\lambda m)P_{0}(m,t)+\lambda(m+1)P_{0}(m+1,t)+g\sum_{\alpha}P_{\alpha}(m,t)
\end{equation}

\begin{equation}
\dot{G}_{0}=-(r+g)G_{0}-\lambda z\partial_{z}G_{0}+\lambda\partial_{z}G_{0}+gG
\end{equation}
{\bf\hspace{0.25in}Steady-state Probability}

\begin{equation}
\dot{G}_{0}=0\Rightarrow G_{0}=\frac{\lambda}{r+g}(1-z)\partial_{z}G_{0}+\frac{g}{r+g}G\Rightarrow P_{0}=\frac{g}{r+g}
\end{equation}

\subsection*{Arrested ($\bm{\alpha=m_c}$)}

\begin{equation}
\dot{P}_{m_c}  =  -(g+\lambda m)P_{m_c}(m,t)+\lambda(m+1)P_{m_c}(m+1,t)+rP_{m_c-1}(m-1,t)
\end{equation}

\begin{equation}
\dot{G}_{m_c}=-gG_{m_c}-\lambda z\partial_{z}G_{m_c}+\lambda\partial_{z}G_{m_c}+zrG_{m_c-1}
\end{equation}

\subsubsection*{\quad Steady-state Probability}

\begin{equation}
\dot{G}_{m_c}=0\Rightarrow G_{m_c}=\frac{\lambda}{g}(1-z)\partial_{z}G_{m_c}+\frac{zr}{g}G_{m_c-1}\Rightarrow P_{m_c}=\frac{r}{g}P_{m_c-1}
\end{equation}

\subsection*{Bulk ($\bm{1 \le \alpha \le m_c-1}$)}

\begin{eqnarray*}
\dot{P}_{\alpha} & = & -(r+g+\lambda m)P_{\alpha}(m,t)+\lambda(m+1)P_{\alpha}(m+1,t)+rP_{\alpha-1}(m-1,t)
\end{eqnarray*}

\begin{equation}
\dot{G}_{\alpha}=-(r+g)G_{\alpha}-\lambda z\partial_{z}G_{\alpha}+\lambda\partial_{z}G_{\alpha}+zrG_{\alpha-1}
\end{equation}

\subsubsection*{\quad Steady-state Probability}

\begin{equation}
\dot{G}_{\alpha}=0\Rightarrow G_{\alpha}=\frac{\lambda}{r+g}(1-z)\partial_{z}G_{m_c}+\frac{zr}{r+g}G_{m_c-1}\Rightarrow P_{\alpha}=\frac{r}{r+g}P_{\alpha-1}
\end{equation}

\subsection{Total Probability }

Iteratively we can get the probabilities for the bulk as 

\begin{equation}
P_{\alpha}=\frac{r}{r+g}P_{\alpha-1}=\frac{r}{r+g}....\frac{r}{r+g}P_{0}=\frac{P_{0}}{(1+\frac{g}{r})^{\alpha}},
\label{Palpha}
\end{equation}
and the arrested state probability has the simple form

\begin{equation}
P_{m_c}=\frac{r}{g}P_{m_c-1}=\frac{1}{(1+\frac{g}{r})^{m_{c}}}
\end{equation}

\subsection{Total Mean }

The equation for  the total generating function equation is obtained by just adding all the equations together to find

\begin{equation}
\dot{G}=\lambda(1-z)\partial_{z}G+r(z-1)\left(G-G_{m_c}\right),
\end{equation}
so the total generating function depends only on the arrested state.
The stationary equation is thus

\begin{equation}
\dot{G}=0\Rightarrow\partial_{z}G=\frac{r}{\lambda}\left(G-G_{m_c}\right)
\end{equation}
Using the standard expression for the mean in terms of the generating
function we find

\begin{equation}
\overline{m}=\partial_{z}G|_{z=1}=\frac{r}{\lambda}\left(G-G_{m_c}\right)|_{z=1}=\frac{r}{\lambda}\left(1 - P_{m_c}\right)=\frac{r}{\lambda}\left(1-\frac{1}{(1+\frac{g}{r})^{m_{c}}}\right)
\end{equation}
It is interesting that $\overline{m}$ goes to a finite limit as $r\to \infty$,

\begin{equation}
\overline{m}\rightarrow\frac{gm_c}{\lambda}
\end{equation}

\subsection{Variance/Fano Factor}

Using the expression for the first moment we can find the second moment

\begin{equation}
\dot{G}=0\Rightarrow\partial_{z}G=\frac{r}{\lambda}\left(G-G_{m_c}\right)\Rightarrow\partial_{z}^{2}G=\frac{r}{\lambda}\left(\partial_{z}G-\partial_{z}G_{m_c}\right)
\end{equation}
This gives

\begin{equation}
\overline{m^{2}}-\overline{m}=\partial_{z}^{2}G|_{z=1}=\frac{r}{\lambda}\left(\partial_{z}G-\partial_{z}G_{m_c}\right)\Big |_{z=1}=\frac{r}{\lambda}\left(\overline{m}-\overline{m}_{m_c}\right)
\end{equation}
where $\overline{m}_{m_c}$ is the mean of $m$ in the arrested state $\alpha=m_c$.
The variance is the given by

\begin{equation}
\overline{m^{2}}-\overline{m}^{2}=(1+\frac{r}{\lambda})\overline{m}-\frac{r}{\lambda}\overline{m}_{m_c}-\overline{m}^{2}
\end{equation}
and the Fano factor by

\[
F=\frac{\overline{m^{2}}-\overline{m}^{2}}{\overline{m}}=(1+\frac{r}{\lambda})-\overline{m}-\frac{r}{\lambda}\frac{\overline{m}_{m_c}}{\overline{m}}=1+\frac{r}{\lambda}\left(\frac{1}{(1+\frac{g}{r})^{m_{c}}}-\frac{\overline{m}_{m_c}}{\overline{m}}\right)
\]
So then we are left with figuring out $\overline{m}_{m_c}$.

\subsubsection*{Relaxed}

\begin{equation}
\dot{G}_{0}=0\Rightarrow(r+g)\partial_{z}G_{0}=\lambda(1-z)\partial_{z}^{2}G_{0}-\lambda\partial_{z}G_{0}+g\partial_{z}G
\end{equation}
and so
\begin{equation}
\overline{m}_{0}=\frac{g}{r+\lambda+g}\overline{m}
\end{equation}

\subsubsection*{Arrested}

\begin{equation}
\dot{G}_{m_c}=0\Rightarrow\partial_{z}G_{m_c}=-\frac{\lambda}{g}\partial_{z}G_{m_c}+\frac{zr}{g}\partial_{z}G_{a-1}+\frac{r}{g}G_{a-1}
\end{equation}
giving
\begin{equation}
\overline{m}_{m_c}=\frac{r}{g+\lambda}(\overline{m}_{m_c-1}+P_{m_c-1})=B(\overline{m}_{m_c-1}+P_{m_c-1})
\label{eqmmc}
\end{equation}
where
\begin{equation}
B\equiv \frac{r}{g+\lambda}
\end{equation}

\subsubsection*{Bulk}

\begin{equation}
\dot{G}_{\alpha}=0\Rightarrow(r+g)\partial_{z}G_{\alpha}=\lambda(1-z)\partial_{z}^{2}G_{0}-\lambda\partial_{z}G_{m_c}+zr\partial_{z}G_{\alpha-1}+rG_{\alpha-1}
\end{equation}
implying
\begin{equation}
\overline{m}_{\alpha}=\frac{r}{r+g+\lambda}(\overline{m}_{\alpha-1}+P_{\alpha-1})=A(\overline{m}_{\alpha-1}+P_{\alpha-1})
\label{eqmalf}
\end{equation}
where
\begin{equation}
A\equiv \frac{r}{r+g+\lambda}
\end{equation}

\subsubsection*{Solution}

This is a bit complicated due to the recursive nature of the mean.
First we must write the bulk mean contributions to the total mean
in terms of the first state. 
The inhomogeneous linear recursion relation Eq. \ref{eqmalf}
has the solution
\begin{equation}
m_{\alpha}  =  A^{\alpha}m_{0}+\sum_{\beta=0}^{\alpha-1}A^{\alpha-\beta}P_{\beta}
\end{equation}
Then, by Eq. \ref{eqmmc},

\begin{align}
m_{m_c}&=BA^{m_{c}-1}m_{0}+BA^{m_{c}-1}\sum_{\beta=0}^{m_{c}-2}A^{-\beta}P_{\beta}+BP_{m_c-1}\nonumber\\
&=BA^{m_{c}-1}m_{0}+BA^{m_{c}-1}\sum_{\alpha=0}^{m_{c}-1}A^{-i}P_{\alpha}
\end{align}

We can do the sum since, from Eq.  \ref{Palpha}, the $P_\alpha$ are geometric,
and so

\begin{equation}
m_{m_c}=BA^{m_{c}-1}\frac{g}{r+\lambda+g}\overline{m}+BA^{m_{c}-1}P_{0}\frac{(\frac{P}{A})^{m_{c}}-1}{\frac{P}{A}-1}
\end{equation}

\begin{figure}
\begin{center}
\includegraphics[width=0.9\textwidth]{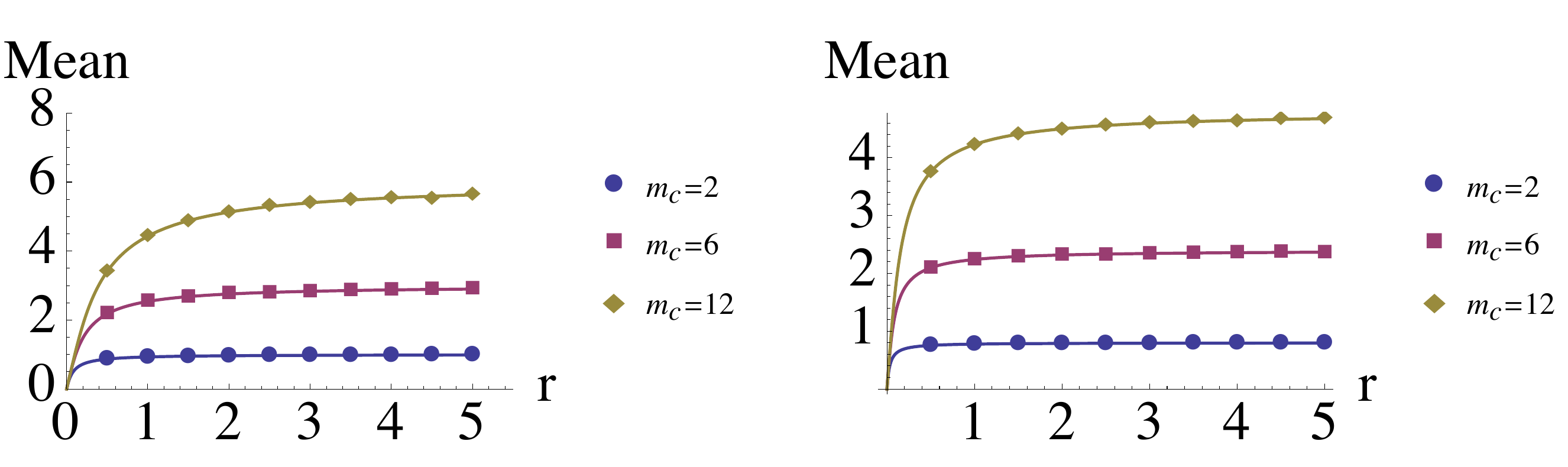}\\
\hspace*{3pt}\includegraphics[width=0.9\textwidth]{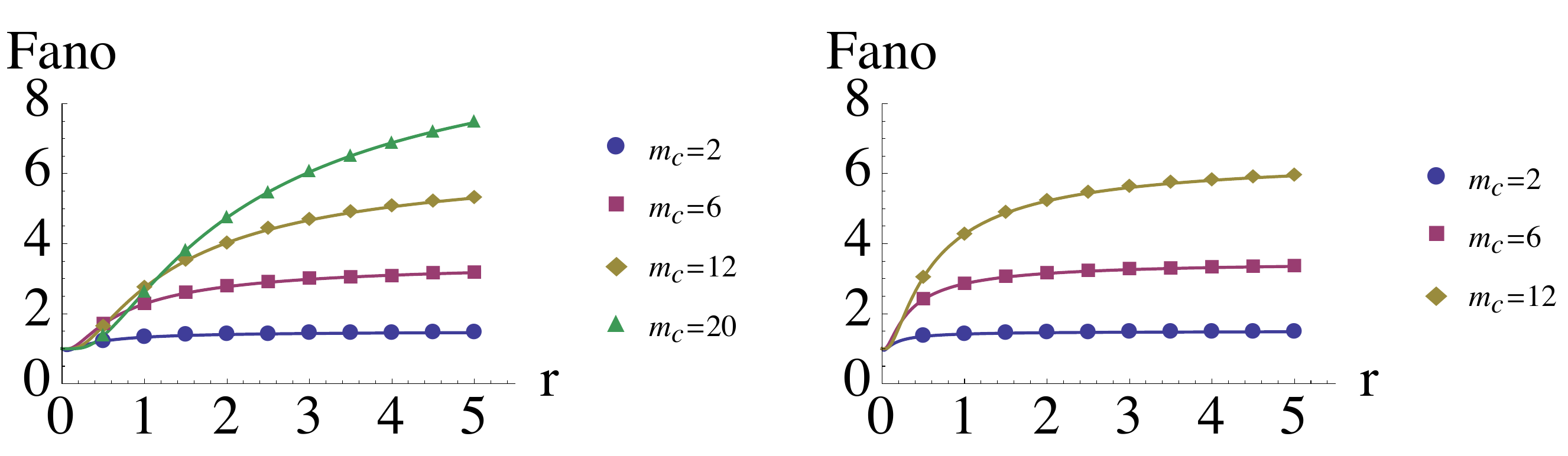}
\end{center}
\caption{ The left panel is for a gene with fixed $\lambda=\frac{1}{20},\;g=\frac{1}{20}$
and the right for a gene with fixed $\lambda=\frac{1}{20},\;g=\frac{1}{50}$
for various cutoff values $m_{c}$. Solid curves are analytical expressions
and dots numerical data. The numerical results were obtained through
standard Gillespie algorithm methods using the mechanically based
regulation outlined above. }
\label{fig3}
\end{figure}

The analytical expressions are compared to numerical simulations in Fig. \ref{fig3}.

\subsection{Curve Scaling}

We are especially interested in obtaining the functional form for
the Fano factor mean curves when $F>1$. To do this we will explore
the large $r,\ m_{c}$ limit for fixed $\frac{r}{m_{c}}$. In this
limit the mean becomes 

\begin{equation}
\overline{m}=\frac{r}{\lambda}\left(1-e^{-g\frac{m_{c}}{r}}\right)
\end{equation}
For $\frac{r}{g m_{c}}\ll 1$, $\overline{m}\sim\frac{r}{\lambda}$
which matches the well known form for simple random mRNA production
and degradation. On the other hand, for large $\frac{r}{g m_{c}}\rightarrow\overline{m}\sim m_{c}\frac{g}{\lambda}$.
Similarly, for the variance 

\begin{equation}
\overline{m^{2}}-\overline{m}^{2}=\frac{r^2e^{-m_{c}\frac{g}{r}}}{\lambda^{2}\left(g+\lambda\right)}\left[\lambda\left(1-e^{-m_{c}\frac{g}{r}}\right)-ge^{-m_{c}\frac{g}{r}}\left(1-e^{-m_{c}\frac{\lambda}{r}}\right)\right]
\end{equation}
These expressions become more clear in particular cases such as for
$g=\lambda$ where we can express the mean and Fano factor  as 

\begin{equation}
\overline{m}\sim m_{c}x\left(1-e^{-\frac{1}{x}}\right),\;F\sim m_{c}\frac{xe^{-\frac{1}{x}}\left(1-e^{-\frac{1}{x}}\right)}{2}=\frac{\overline{m}e^{-1/x}}{2},\quad x\equiv\frac{r}{m_{c}g}
\end{equation}
We see that $F$  is exponentially
small for small $x$ and approaches $\frac{\overline{m}}{2}$ for $x\gg1$.
Thus, the Fano factor has a crossover from being very small for small
$x$ and becomes order $1$ for large $x$ . 

For $\lambda=2g$, we get

\begin{equation}
F\sim m_{c}\frac{xe^{-\frac{1}{x}}}{6}\left(1-e^{-\frac{1}{x}}\right)\left(2+e^{-\frac{1}{x}}\right) = \frac{\overline{m}e^{-1/x}}{3}\left(2+e^{-\frac{1}{x}}\right)
\end{equation}
so that at small $x$ it is roughly a factor at $4/3$ larger than the $\lambda=g$ case, increasing to a factor of 2 for the same $\overline{m}$. Nevertheless,  for large $x$ it obviously reproduces the universal limit $(m_c+1)/2$
to leading order in $m_c$, since the saturated value of $\overline{m}$ is a factor of 2 smaller.

Lastly for $\lambda=3g$, we get

\begin{equation}
F\sim m_{c}\frac{xe^{-\frac{1}{x}}}{12}\left(1-e^{-\frac{1}{x}}\right)\left(3+2e^{-\frac{1}{x}}+e^{-\frac{2}{x}}\right) =  \frac{\overline{m}e^{-1/x}}{4}\left(3+2e^{-\frac{1}{x}}+e^{-\frac{2}{x}}\right)
\end{equation}
which is larger still for small $x$ and approaches the universal limit for large $x$.

\subsection{Slowed Transcription}

 Within the basic phenomenological model outlined here there are a number of simplifications which we have employed that are worth addressing, most notably the nature of the mechanical arrest and relaxation processes.  In our simplified model the production of any given mRNA adds the same mechanical strain to the system and each occurs with the same random rate  $r$ up to the cutoff $m_{c}$.  This choice has support from experimental data \cite{Golding2005} showing that the time to transcribe a burst of $N$ transcripts is the same as the time to transcribe $N$ transcripts independently.   Additionally, the RNAP machinery velocity remains fairly constant as it approaches its maximum operating load \cite{Ma2013}.  Nonetheless, it is certainly worth considering a more nuanced view of the arrest process in which there is a slowing associated with RNA production in a strained but not stalled gene.  We have conducted numerical simulations (shown in Figure 7 as points) for a gene which experiences decreased proclivity for expression as the cutoff $m_{c}$ is approached through a  decreased mRNA production rate $r_\textit{eff}=r/(1+\alpha)$ where $r$ is the baseline rate and $\alpha$ is the number of transcripts made since the last relaxation event.   The introduction of this slowing effect does not lead to qualitative changes in our findings; only minor quantitative differences are seen in the numerical results as compared to the analytical expressions from the simpler model.

 \begin{figure}[slow]
\begin{center}
\includegraphics[width=0.6\textwidth]{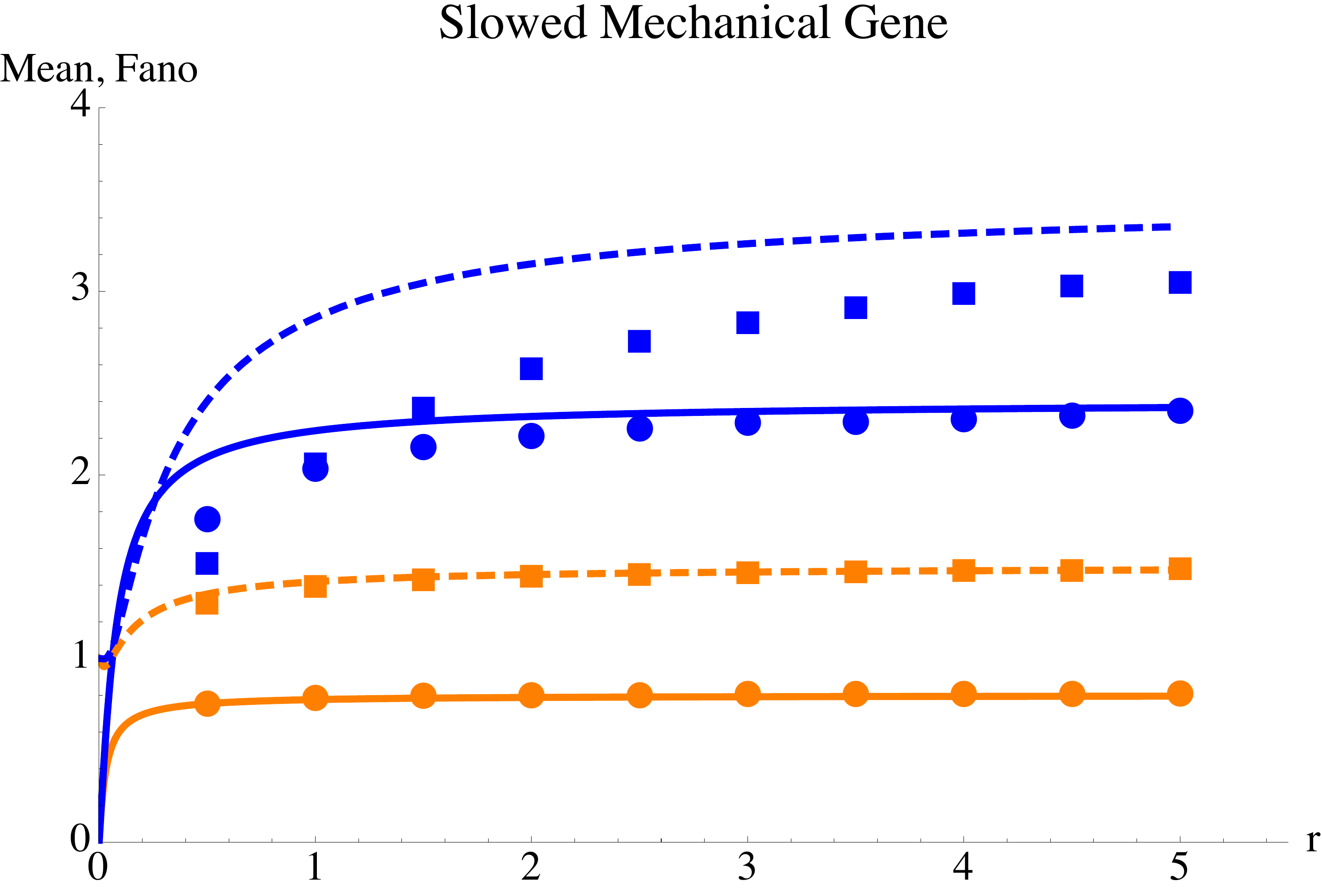}
\end{center}
\caption{(color online) Analytical curves showing the mean mRNA (solid, circles) and Fano factor (dashed, squares) with no slowing against numerical data for the extended model in which transcription slows as the cutoff is approached as described in letter for various cutoffs $m_{c}=2,6$ (orange, blue) at $\lambda=\frac{1}{50},g=\frac{1}{20} $.}
\label{fig2}
\end{figure}

\section{Model with a Repressor: Generating Function Approach}

We now repeat the same formulation but incorporate a suppressed
state by constructing a generating function for each state which is
unsuppressed $G_{\alpha}(z,t)=\sum_{m}z^{m}P_{\alpha}(m,t)$ and suppressed $H_{\alpha}(z,t)=\sum_{m}z^{m}Q_{\alpha}(m,t)$
each of which corresponds again to a given DNA mechanical state $\alpha$. Again we
will write the master equations for each distinctive state type.

\subsection{Equations}

\subsection*{Relaxed}

\begin{align}
\dot{P}_{0} & =  -(r+g+k_{+}+\lambda m)P_{0}(m,t)+\lambda(m+1)P_{0}(m+1,t)+g\sum_{\alpha}P_{\alpha}(m,t)+k_{-}Q_{0}\nonumber\\
\dot{Q}_{0} & =  -(g+k_{-}+\lambda m)Q_{0}(m,t)+\lambda(m+1)Q_{0}(m+1,t)+g\sum_{\alpha}Q_{\alpha}(m,t)+k_{+}P_{0}
\end{align}

\begin{align}
\dot{G}_{0} &=-(r+g+k_{+})G_{0}+\lambda(1-z)\partial_{z}G_{0}+k_{-}H_{0}+gG \nonumber\\
\dot{H}_{0} &=-(g+k_{-})H_{0}+\lambda(1-z)\partial_{z}H_{0}+k_{+}G_{0}+gH
\end{align}

\subsection*{Arrested}

\begin{align}
\dot{P}_{m_c} & =  -(g+k_{+}+\lambda m)P_{m_c}(m,t)+\lambda(m+1)P_{m_c}(m+1,t)+rP_{m_c-1}(m-1,t)+k_{-}Q_{m_c}\nonumber\\
\dot{Q}_{m_c} & =  -(g+k_{-}+\lambda m)Q_{m_c}(m,t)+\lambda(m+1)Q_{m_c}(m+1,t)+k_{+}P_{m_c}
\end{align}

\begin{align}
\dot{G}_{m_c} &=-(g+k_{+})G_{m_c}+\lambda(1-z)\partial_{z}G_{m_c}+zrG_{a-1}+k_{-}H_{m_c} \nonumber\\
\dot{H}_{m_c} &=-(g+k_{-})H_{m_c}+\lambda(1-z)\partial_{z}H_{m_c}+k_{+}G_{m_c}
\label{eqGHmc}
\end{align}

\subsection*{Bulk}

\begin{align}
\dot{P}_{\alpha} & =  -(r+g+k_{+}+\lambda m)P_{\alpha}(m,t)+\lambda(m+1)P_{\alpha}(m+1,t)+rP_{\alpha-1}(m-1,t)+k_{-}Q_{\alpha}\nonumber\\
\dot{Q}_{\alpha} & =  -(g+k_{-}+\lambda m)Q_{\alpha}(m,t)+\lambda(m+1)Q_{\alpha}(m+1,t)+k_{+}P_{\alpha}
\end{align}

\begin{align}
\dot{G}_{\alpha}&=-(r+g+k_{+})G_{\alpha}+\lambda(1-z)\partial_{z}G_{\alpha}+zrG_{\alpha-1}+k_{-}H_{\alpha} \nonumber\\
\dot{H}_{\alpha}&=-(g+k_{-})H_{\alpha}+\lambda(1-z)\partial_{z}H_{\alpha}+k_{+}G_{\alpha}
\end{align}

\subsubsection*{Together}

\begin{align}
\dot{G}&=-r(G-G_{f})+\lambda(1-z)\partial_{z}G+zr(G-G_{f})-k_{+}G+k_{-}H \nonumber\\
\dot{H} &=-k_{-}H+\lambda(1-z)\partial_{z}H+k_{+}G
\end{align}
Then

\begin{equation}
\dot{H}+\dot{G}=\lambda(1-z)\partial_{z}(H+G)+r(z-1)(G-G_{m_c})
\end{equation}
so that 

\begin{equation}
\partial_{z}(H+G)=\frac{r}{\lambda}(G-G_{m_c})
\end{equation}
and so
\begin{align}
\overline{m}&=\frac{r}{\lambda}(P-P_{m_c}),\nonumber\\
\overline{m^{2}}-\overline{m}&=\frac{r}{\lambda}\left(\overline{m}^G-\overline{m}^G_{m_c}\right)
\label{eqmvar}
\end{align}
where $\overline{m}^G$ is the mean number of mRNAs over all unsuppressed states and $\overline{m}^G_{m_c}$ is the mean number of mRNAs in the unsuppressed, arrested state.

\subsection{Mean }

Finding the probability $P$ of being unsuppressed is rather easy given that
$P=\frac{k_{-}}{k_{+}}Q,\;P+Q=1$, so that
\begin{equation}
 P=\frac{k_{-}}{k_{+}+k_{-}};\ Q=\frac{k_{+}}{k_{+}+k_{-}}.
\end{equation}

The result for the unsuppressed, arrested state $P_{m_c}$ is a little harder but
we will first start by seeing for all the $Q$ states except for the
relaxed state $Q_{0}$ we have

\begin{equation}
Q_{\alpha}=\frac{k_{+}}{k_{-}+g}P_{\alpha},\;\alpha\neq0
\end{equation}
We can now tackle the $P$ states on their own

\begin{align}
P_{m_c}=&\frac{r}{g+k_{+}}P_{m_c-1}+\frac{k_{-}}{g+k_{+}}Q_{m_c}=\frac{r}{g+k_{+}}P_{m_c-1}+\frac{k_{-}}{g+k_{+}}\frac{k_{+}}{k_{-}+g}P_{m_c}\nonumber\\
&=\left[\frac{r}{g+k_{+}-k_{-}k_{+}/(g+k_{-})}\right]P_{a-1}\equiv AP_{a-1}
\end{align}
where
\begin{equation}
A\equiv\frac{r}{g+k_{+}-k_{-}k_{+}/(g+k_{-})}
\end{equation}
Similarly, for the bulk states,

\begin{equation}
P_{\alpha}=\left[\frac{r}{r+g+k_{+}-k_{-}k_{+}/(g+k_{-})}\right] P_{\alpha-1}\equiv BP_{\alpha-1}
\label{Palf1}
\end{equation}
where
\begin{equation}
B\equiv \frac{r}{r+g+k_{+}-k_{-}k_{+}/(g+k_{-})}
\end{equation}
For the relaxed state 

\begin{align}
(r+g+k_{+})P_{0}&=gP+k_{-}Q_{0}\nonumber\\
Q_{0}& =\frac{g}{k_{-}+g}Q+\frac{k_{+}}{k_{-}+g}P_{0}=\frac{g}{k_{-}+g}\frac{k_{+}}{k_{-}}P+\frac{k_{+}}{k_{-}+g}P_{0}
\end{align}
Then

\begin{equation}
P_{0}=\left(r+g+k_{+}-\frac{k_{+}k_{-}}{k_{-}+g}\right)^{-1}\left(g+g\left(\frac{k_{-}}{k_{-}+g}\right)\frac{k_{+}}{k_{-}}\right)P\equiv CP
\label{P01}
\end{equation}
where
\begin{align}
C &\equiv\left(r+g+k_{+}-\frac{k_{+}k_{-}}{k_{-}+g}\right)^{-1}\left(g+g\left(\frac{k_{-}}{k_{-}+g}\right)\frac{k_{+}}{k_{-}}\right)\nonumber\\
&= \left(\frac{B}{r}\right) \frac{g(k_-+k_+ + g)}{k_-+g} = \left(\frac{B}{r}\right)\left(\frac{r}{A}\right) = \frac{B}{A}
\label{eqC}
\end{align}
Iterating Eq. (\ref{Palf1}) together with  Eqs. (\ref{P01},\ref{eqC}) , we have 

\begin{equation}
P_{m_c}=AB^{m_{c}-1}CP = B^{m_c}P
\end{equation}
yielding a mean mRNA level expression 

\begin{align}
\overline{m} & =  \frac{r}{\lambda}P(1-B^{m_{c}})\nonumber\\
 & =  \frac{r}{\lambda}\cdot\frac{k_{-}}{k_{+}+k_{-}}\cdot\left(1-\left(\frac{r}{r+g+g\frac{k_{+}}{g+k_{+}}}\right)^{m_{c}}\right)
\end{align}

\subsection{Fano Factor}
We now turn to the calculation of $\overline{m}^G$ and $\overline{m}^G_{m_c}$.

\subsubsection*{Arrested}

Starting from Eqs. (\ref{eqGHmc}),
setting the time derivatives to zero  and taking $z\rightarrow 1$ after the derivatives
are taken yields

\begin{align}
(g+k_{+}+\lambda)\partial_{z}G_{m_c} &=r\left(G_{m_c-1}+\partial_{z}G_{m_c-1}\right)+k_{-}\partial_{z}H_{m_c} \nonumber\\
(g+k_{-}+\lambda)\partial_{z}H_{m_c} &=k_{+}\partial_{z}G_{m_c}
\label{eqmghmc}
\end{align}

This gives

\begin{equation}
m^G_{m_c}=\frac{r}{g+k_{+}+\lambda-\frac{k_{-}k_{+}}{g+k_{-}+\lambda}}\left(m^G_{m_c-1}+P_{m_c-1}\right)\equiv F\left(m^G_{m_c-1}+P_{m_c-1}\right)
\label{eqmgmc}
\end{equation}
where
\begin{equation}
F \equiv \frac{r}{g+k_{+}+\lambda-\frac{k_{-}k_{+}}{g+k_{-}+\lambda}}\
\end{equation}

\subsubsection*{Bulk}

\begin{align}
(r+g+k_{+}+\lambda)\partial_{z}G_{\alpha}&=r\left(G_{\alpha-1}+\partial_{z}G_{\alpha-1}\right)+k_{-}\partial_{z}H_{\alpha}\nonumber\\
(g+k_{-}+\lambda)\partial_{z}H_{\alpha}&=k_{+}\partial_{z}G_{\alpha}
\label{eqmghalf}
\end{align}
giving

\begin{equation}
m^G_{\alpha}=\frac{r}{r+g+k_{+}+\lambda-\frac{k_{-}k_{+}}{g+k_{-}+\lambda}}\left(m_{\alpha-1}+P_{\alpha-1}\right)\equiv Y\left(m^G_{\alpha-1}+P_{\alpha-1}\right)
\label{eqmgalf}
\end{equation}
where
\begin{equation}
Y\equiv \frac{r}{r+g+k_{+}+\lambda-\frac{k_{-}k_{+}}{g+k_{-}+\lambda}}
\end{equation}

\subsubsection*{Relaxed}

\begin{align}
(r+g+k_{+}+\lambda)\partial_{z}G_{0}&=k_{-}\partial_{z}H_{0}+g\partial_{z}G \nonumber\\
(g+k_{-}+\lambda)\partial_{z}H_{0}&=k_{+}\partial_{z}G_{0}+g\partial_{z}H
\label{eqmgh0}
\end{align}
so that

\begin{equation}
m^G_{0}=\frac{g}{r+g+k_{+}+\lambda-\frac{k_{-}k_{+}}{g+k_{-}+\lambda}}\left(\overline{m}^G+\frac{k_{-}}{g+k_{-}+\lambda}\overline{m}^H\right) \nonumber\\
\end{equation}
We know $\overline{m}$, but we still need $\overline{m}^G$ and $\overline{m}^H$.  To find these, we sum the
second lines of Eqs. (\ref{eqmghmc}, \ref{eqmghalf}, \ref{eqmgh0}) and find
\begin{equation}
 \overline{m}^H=\frac{k_{+}}{\lambda+k_{-}}\overline{m}^G
\end{equation}
Then, using $\overline{m}^G+\overline{m}^H=\overline{m}$, we get
\begin{equation}
\overline{m}^G = \frac{\lambda + k_-}{\lambda+k_- + k_+}\overline{m}; \qquad \overline{m}^H = \frac{ k_+}{\lambda+k_- + k_+}\overline{m}
\end{equation}
and
\begin{equation}
m_{0}=\frac{g}{r+g+k_{+}+\lambda-\frac{k_{-}k_{+}}{g+k_{-}+\lambda}}\left(1+\frac{k_{-}}{g+k_{-}+\lambda}\frac{k_{+}}{\lambda+k_{-}}\right)m^G\equiv Vm^G
\end{equation}
where 
\begin{equation}
V \equiv \frac{g}{r+g+k_{+}+\lambda-\frac{k_{-}k_{+}}{g+k_{-}+\lambda}}\left(1+\frac{k_{-}}{g+k_{-}+\lambda}\frac{k_{+}}{\lambda+k_{-}}\right)
\end{equation}
Solving the linear recursion Eq. (\ref{eqmgalf}), we have
\begin{equation}
m^G_\alpha = Y^\alpha + \sum_{\beta=0}^{\alpha-1} Y^{\alpha-1-\beta} P_\beta
\end{equation}
so from Eq. (\ref{eqmgmc}
\begin{align}
m_{m_c}&=FY^{m_{c}-1}m_{0}+FY^{m_{c}-1}\sum_{\alpha=0}^{m_{c}-2}Y^{-\alpha}P_{\alpha}+FP_{m_c-1}\nonumber\\
&=FY^{m_{c}-1}m_{0}+FY^{m_{c}-1}\sum_{\alpha=0}^{m_{c}-1}Y^{-\alpha}P_{\alpha}
\end{align}
which can be computing since $P_\alpha$ obey a geometric distribution, Eq. (\ref{Palf1}).

The analytical expressions are compared to numerical simulations below.
All results are for a gene with fixed $\lambda=\frac{1}{10},\;g=\frac{1}{10},\;a,m_{c}=5,\;k_{-}=\frac{1}{50}$. 

\begin{figure}
\begin{center}
\includegraphics[scale=0.6]{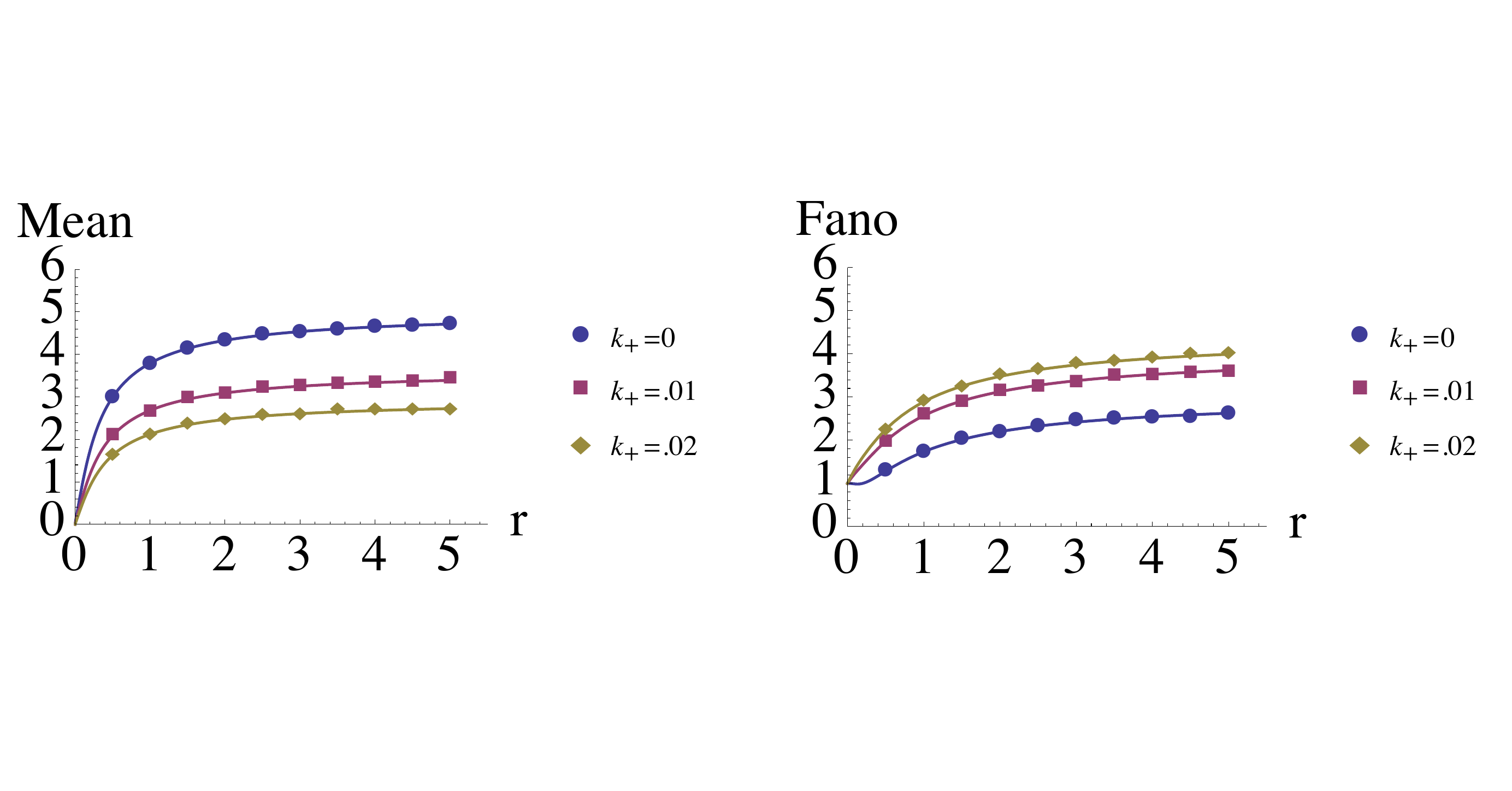}
\end{center}
\caption{
Numerical results obtained through standard Gillespie algorithm
methods using the mechanically based regulatory network outlined above
compared to analytical results. }
\label{fig4}
\end{figure}

\subsection{Limits}

In the fast transcriptional limit for a gene with repression the $r$
independent Fano factor becomes 

\begin{equation}
F\stackrel[r\to\infty]{\xrightarrow{1.5em}}{}\frac{1+m_{c}}{2}+m_{c}\left(\frac{g^{2}k_{+}(g+k_{-}+\lambda)-gk_{+}k_{-}(k_{+}+k_{-})}{(k_{+}+k_{-})(k_{+}+k_{-}+\lambda)(k_{-}+g+\lambda)}\right)
\end{equation}
In the limit of $k_{\pm}\ll g,\lambda$ this becomes

\begin{figure}[!h]
\begin{center}
\includegraphics[scale=0.3]{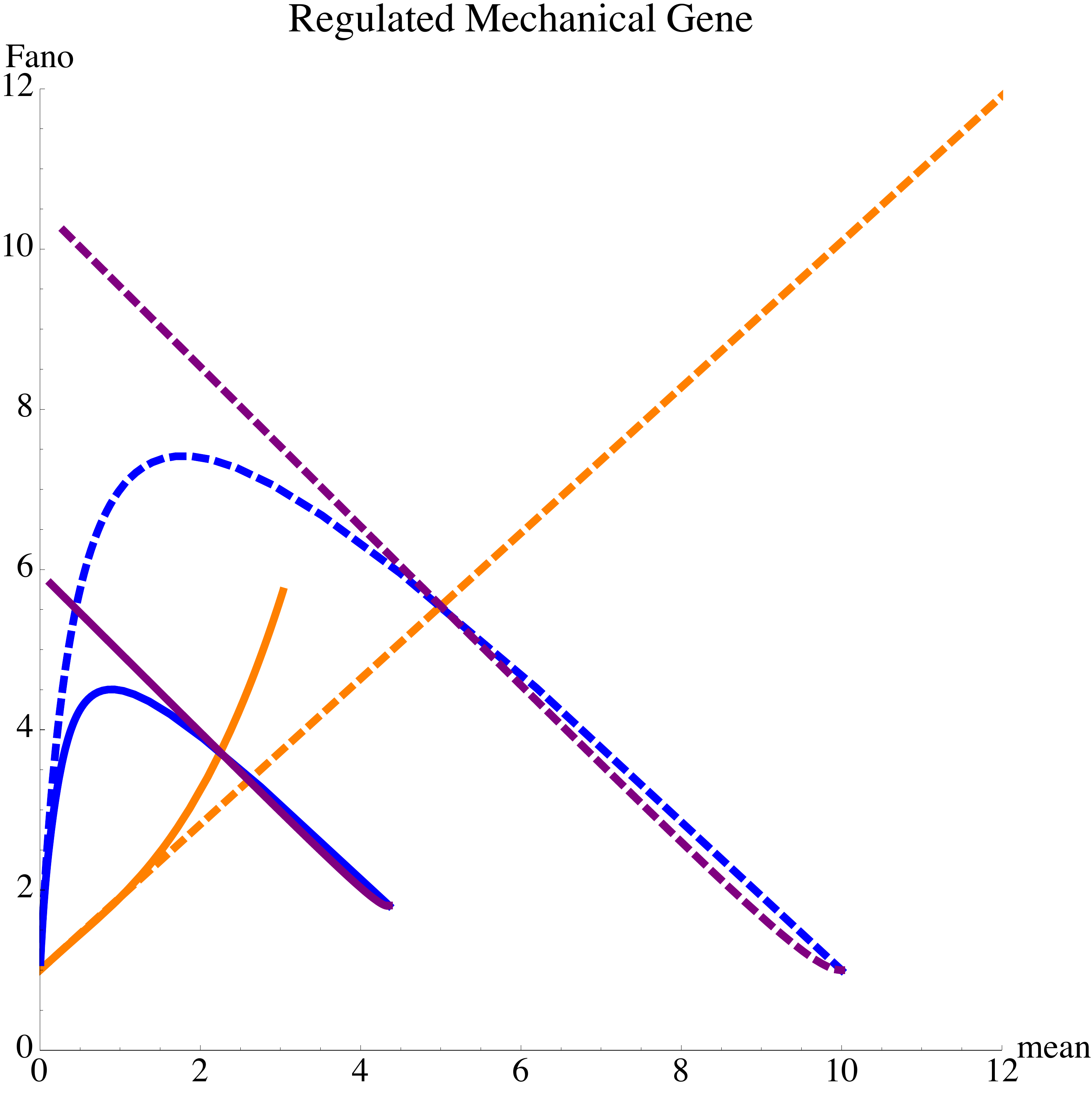}
\end{center}
\caption{
Each solid curve corresponds to varying a single parameter
$r$ (orange), repressor binding $k_{+}$(blue) and unbinding $k_{-}$(purple)
while holding the others fixed with mechanical limitation to transcription
(with $m_{c}=6,\;\lambda=\frac{1}{20}$ ). The dashed lines show the
same without mechanical governance ($m_{c}\rightarrow\infty)$. It
is interesting to note that the incorporation of the mechanical limitation
does not qualitatively alter the repressor binding/unbinding curves
but instead diminished the overall level of expression and noise.
The promoter strength $r$ parameterized behavior for a mechanically
limited gene is however qualitatively changed showing a non-linear increased
noise/mean relationship compared to the gene with a repressor only. }
\label{fig5}
\end{figure}

\begin{equation}
F\rightarrow\frac{1+m_{c}}{2}+m_{c}g\frac{k_{+}}{\lambda(k_{+}+k_{-})}
\end{equation}
while in the limit $\lambda\ll k_{+},\;k_{-}\ll g$ the Fano factor
is 

\begin{equation}
F\rightarrow\frac{1+m_{c}}{2}+m_{c}g\frac{k_{+}}{(k_{+}+k_{-})^{2}}
\end{equation}
This illustrates how noise much greater than the lower bound set by the
mechanical limit can exist for various repressor kinetics.

In Fig. \ref{fig5} we show curves demonstrating mean/Fano relationships for varying binding kinetics
$k_{\pm}$.

\section{Wait-Time Distributions}
Here we calculate the wait-time distributions for the probability of taking time $t$ to reach the arrested state from the relaxed state.  The problem is taken as a first passage time problem with an absorbing boundary at the arrested state. 
\begin{equation}
P_{on\rightarrow off}(t)=-\partial_{t}P_{on},\;P_{on}=\sum_{i=0}^{m_c-1}P_{i}
\end{equation}

We'll only need the transitions between the states and have no need
to incorporate RNA production/degradation thus we'll be interested
in the master equation dynamics of the states by themselves

\begin{align}
\partial_{t}P_{0} & =  -(r+g+k)P_{o}+g\nonumber\\
\partial_{t}P_{i} & =  -(r+g+k)P_{i}+rP_{i-1}
\end{align}

Taking the Laplace transform yields 

\begin{equation}
\tilde{P}_{0}(s)=\frac{g+1}{s+r+g+k},\;\tilde{P}_{i}(s)=\frac{r}{s+r+g+k}\tilde{P}_{i-1}(s)
\end{equation}

Adding these all together yields

\begin{equation}
\tilde{P}_{on}(s)=\sum_{i=0}^{m_c-1}\tilde{P}_{i}(s)=\tilde{P}_{0}\sum_{n=0}^{m_c-1}\left(\frac{r}{s+r+g+k}\right)^n
\end{equation}

Thus

\begin{eqnarray*}
P_{on}(t)&=&\mathcal{L}^{-1}\left\{\frac{1+g}{s+g+k}\left(1-\left(\frac{r}{s+r+k}\right)^{m_{c}}\right)\right\} \\
\end{eqnarray*}

These results are used to generate wait-time probability distributions in figure 4a of the main text.

\end{document}